# SYNERGISTIC EFFECT IN HYBRID PLASMONIC CONJUGATES FOR THERANOSTIC APPLICATIONS

*Viktoriia Savchuk[1,2], Ruizheng Wang[3], Lyle Small[3], and Anatoliy Pinchuk[1,2]*

[1]Department of Physics and Energy Science, University of Colorado Colorado Springs, 1420 Austin Bluffs Parkway, Colorado Springs, CO 80918, USA

[2]Biofrontiers Institute and Department of Physics and Energy Science, University of Colorado Colorado Springs, 1420 Austin Bluffs Pkwy, Colorado Springs, Colorado 80918, USA

[3]CTI - Chromatic Technologies, Inc., 1096 Elkton Dr., Colorado Springs, CO 80907

## Abstract

Photothermal conversion efficiency ($\eta$) plays a crucial role in selecting suitable gold nanoparticles for photothermal therapeutic applications. The photothermal efficiency depends on the material used for the nanoparticles as well as their various parameters such as size and shape. By maximizing the light-to-heat conversion efficiency ($\eta$), one can reduce the concentration of nanoparticle drugs for photothermal cancer treatment and apply lower laser power to irradiate the tumor. In our study, we explored a new hybrid plasmonic conjugate for theranostic (therapy+diagnostic) applications. We conjugated 20 nm gold nanospheres with cyanine IR dyes, which exhibited significantly enhanced photothermal properties compared to bare nanoparticles. Moreover, the improved photothermal properties of the conjugates can be explained by the synergistic effect that results from the coupling between the metal nanosphere and organic dye.

## 1.1 Introduction

Recent progress in nanotechnology has shown great promise in using plasmonic nanoparticles for Photothermal Therapy (PTT) as a non-invasive type of cancer treatment that is locally applied to the tumor [1-4]. PTT is based on the injection of metal nanoparticles directly into the tumor followed by irradiation of the tumor with a laser. If the wavelength of the laser is matched with localized surface plasmon resonance (LSPR) of nanoparticles, then the incident light can be effectively absorbed. Once absorbed by the nanoparticles, the light energy can be converted to thermal energy, [5-9] leading to necrosis of the targeted cancer cells [10-14].

Gold nanoparticles are a promising candidate for PTT due to their low cytotoxicity and excellent biocompatibility [15]. Their light-to-heat photo-conversion efficiency depends on the wavelengths of both the LSPR and the incident light [16].

One of the most desirable tasks of nanotechnology, including photothermal therapy for cancer treatment, is developing of multifunction drugs that can both diagnose and selectively treat only malignant cells in the body. The near-infrared (NIR) region provides the spectrum range (650-900 nm) that can be simultaneously used both in imaging and therapeutic applications, because the main components of the tissue have the lowest absorption coefficient in this range of wavelengths. [17-19] In addition, the LSPR of metal nanoparticles can be tuned to the desired wavelength in the NIR region to achieve the maximum absorption cross-section. There are several papers that have studied photothermal properties of different shapes and sizes of gold nanoparticles, such as nanospheres, nano-urchins, nanorods, nano-shells, etc., where their ability to produce sufficient heat for cancer treatment has been repeatedly demonstrated. [20-24] Most of these studies were designed with passive accumulation in the malignant tumor site that utilizes the enhanced permeability and retention (EPR) effect [25]. However, passive targeting of metal nanoparticles may and often does result in their accumulation in healthy tissues (e.g., the liver) and can lead to severe side effects. To enhance the uptake of nanoparticles by cancer cells, one can take advantage of active targeting by using gold nanoparticles conjugated with antibodies [10, 26-27]. However, recent studies have shown that the antibody-conjugated gold nanoparticles might be not the only solution for "active targeted drug delivery". Moreover, antibody-conjugated particles don't improve the photothermal properties of gold particles in general. Using infrared organic dyes as an alternative for active cancer targeting may be advantageous not only for near-infrared imaging of cancer [28-30], but also it may increase the uptake of nanoparticles by the cancer cells [31, 32] and potentially improve the efficacy of NIR laser treatment.

In this paper, we introduce a new hybrid plasmonic conjugated particle that can be used for theranostic (therapy+diagnostic) applications of cancer treatment. We present experimental results on the photothermal properties of spherical 20 nm gold nanoparticles (20 nm AuNPs) and 20 nm AuNPs conjugated with an organic tumor-targeting cyanine IR dye (AuNP conjugates) irradiated with a continuous wave of NIR laser ($\lambda_{in} = 808\ nm$). Here, we demonstrate that AuNP conjugates have LSPR in the visible spectral region around 520 nm, and a second absorption peak close to the laser wavelength ($\lambda_{in} = 808\ nm$) that improves their photothermal efficiency. Furthermore, we experimentally demonstrate that 20 nm AuNPs conjugated with an infrared cyanine tumor-targeting dye exhibit a synergistic effect that leads to at least four times higher light-to-heat conversion efficiency as compared to bare 20 nm AuNPs. The enhanced photothermal properties of the AuNP conjugates is beneficial for their applications in PTT since it might provide the required temperature for the PTT treatment with lower concentrations of gold nanoparticles, which may also decrease the unwanted distribution of gold nanoparticles throughout the body as well as with lower incident power of the laser.

## 1.2 Experimental Materials and Experimental Setup

The 20 nm AuNPs in ultra-purified water (UPW) (resistivity ~18 $M\Omega$) were purchased from Ted Pella (Redding, CA). The LSPR peak of the colloidal solution is around 520 nm (Fig. 1(a)). The cyanine IR dyes that were used to functionalize AuNPs have an absorbance peak at the wavelength 780 nm as shown in Fig. 1(a). Spherical AuNP conjugates in UPW were supplied by Lahjavida (Colorado Springs, CO), where 20 nm AuNPs were purchased from Nanopartz (Loveland, CO). The colloidal solutions of the conjugates have two absorbance peaks around 520 nm and 780 nm as indicated in Fig. 1(a). The first peak is associated with LSPR of gold nanosphere, and the second peak is due to the synthesis of these nanoparticles with near-infrared dyes. The cyanine IR dyes are attached to the surface of gold nanoparticles with a linker molecule.

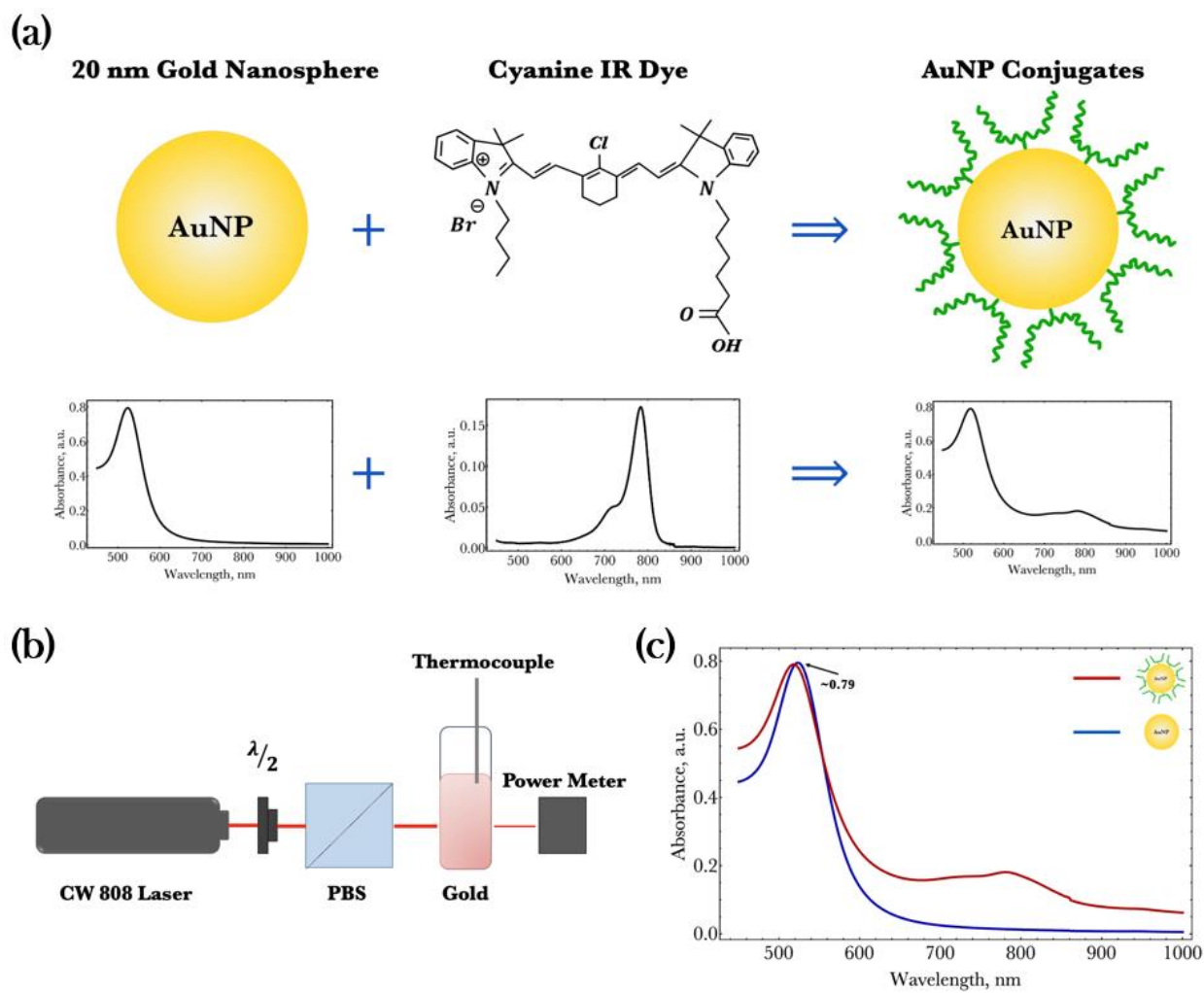

**Figure 1.** (a) Schematic representation and absorbance spectrum of 20 nm Gold Nanosphere (20 nm AuNP) in UPW, cyanine IR dye in methanol, and AuNP conjugates in UPW. (b) A sketch of the experimental setup for measuring the temperature profiles of different solutions. (c) Absorbance spectra of 20 nm AuNPs and AuNP conjugates in UPW.

Fig. 1 (b) shows the experimental setup that was used to measure the temperature profiles of the solutions under continuous wave (CW) NIR laser irradiation. A NIR laser with a wavelength of $808\ nm$ was used to illuminate the 20 nm AuNPs, AuNP conjugates, and the IR cyanine dye solution. All solutions were prepared in a standard 4.5 ml polystyrene cuvette (Fisherbrand, Pittsburgh, PA 15275). The laser's light has elliptical polarization. A Half-Wave Plate (HWP) was used to linearly polarize the incident light, and Polarizing Beam Splitter (PBS) was used to control the incident power of the laser. The temperature of the solution was recorded in real time by using a temperature sensor (PS-2146, Pasco, Roseville, CA, US) connected to PASCO Capstone software. The temperature measurements were carried out in ambient room temperature (~22 °C) conditions. The transmitted power of the laser was also recorded during the experiment by using a digital power meter (PM100D, Thorlabs, Dachau, Germany).

## 1.2  Energy-Balance Equation

The heating process of gold colloidal solution can be described by using the energy-balance equation [24, 33-35]:

$$m_i C_i \frac{dT}{dt} = Q_{in} - Q_{out}, \qquad (1)$$

where $m_i$ and $C_i$ are mass and specific heat capacity of each component in the solution respectively. $Q_{in}$ is the total energy absorbed by the solution, and $Q_{out}$ is the dissipated energy to the environment.

Since the specific heat capacity of water ($C_W = 4.18 \frac{J}{g*K}$) is larger than the heat capacity of gold ($C_W = 0.129 \frac{J}{g*K}$), and the total mass of AuNPs is much less than water, we can simplify the Eq.(1) as

$$m_W C_W \frac{dT}{dt} = Q_{in} - Q_{out}. \qquad (2)$$

The expressions for absorbed and dissipated energies are given by

$$Q_{in} = \eta(P_0 - P_{tr}), \tag{3}$$

$$Q_{out} = hS(T(t) - T_0), \tag{4}$$

where $\eta$ is the photothermal conversion efficiency, $P_0$ is the incident power of the laser, $P_{tr}$ is the laser power that is transmitted through the solution, $h$ is the heat transfer coefficient, $S$ is the surface are of the cuvette, $T_0$ is the room temperature, and $T(t)$ is the temperature of the solution at a given time.

Using Eq. (3)-(4) we can further simplify Eq. (2) as

$$\frac{d\Delta T}{dt} = \frac{\eta(P_0 - P_{tr})}{m_W C_W} - B\Delta T, \tag{5}$$

where we defined $B = \frac{hS}{m_W C_W}$ as the heat dissipation constant and $\Delta T = T(t) - T_0$.

## 1.3 Experimental Temperature Profiles

In this section we present the experimental temperature profiles for 2.5 ml solutions of 20 nm AuNPs and AuNP conjugates in UPW. We measure $\Delta T$ for three different powers (3 trials for each) of the NIR laser ($\lambda_{in} = 808\ nm$).

The solutions of 20 nm AuNPs and AuNP conjugates in UPW with absorbance ~0.79 around 520 nm were prepared for each trial (Fig. 1 (c)). By matching the LSPR peak, we assume that we have the same concentration of gold spherical nanoparticles in each solution.

Figures 2 (a)-(c) show the temperature profiles for solutions of AuNP conjugates, and Figures 2 (d)-(f) show the temperature profiles for 20 nm AuNPs irradiated by the laser with the power $P_o = 1\,W$, $P_o = 1.5\,W$, and $P_o = 2\,W$. The temperature profiles were recorded for 25 min, and the same time was used to measure the cooling temperature profiles.

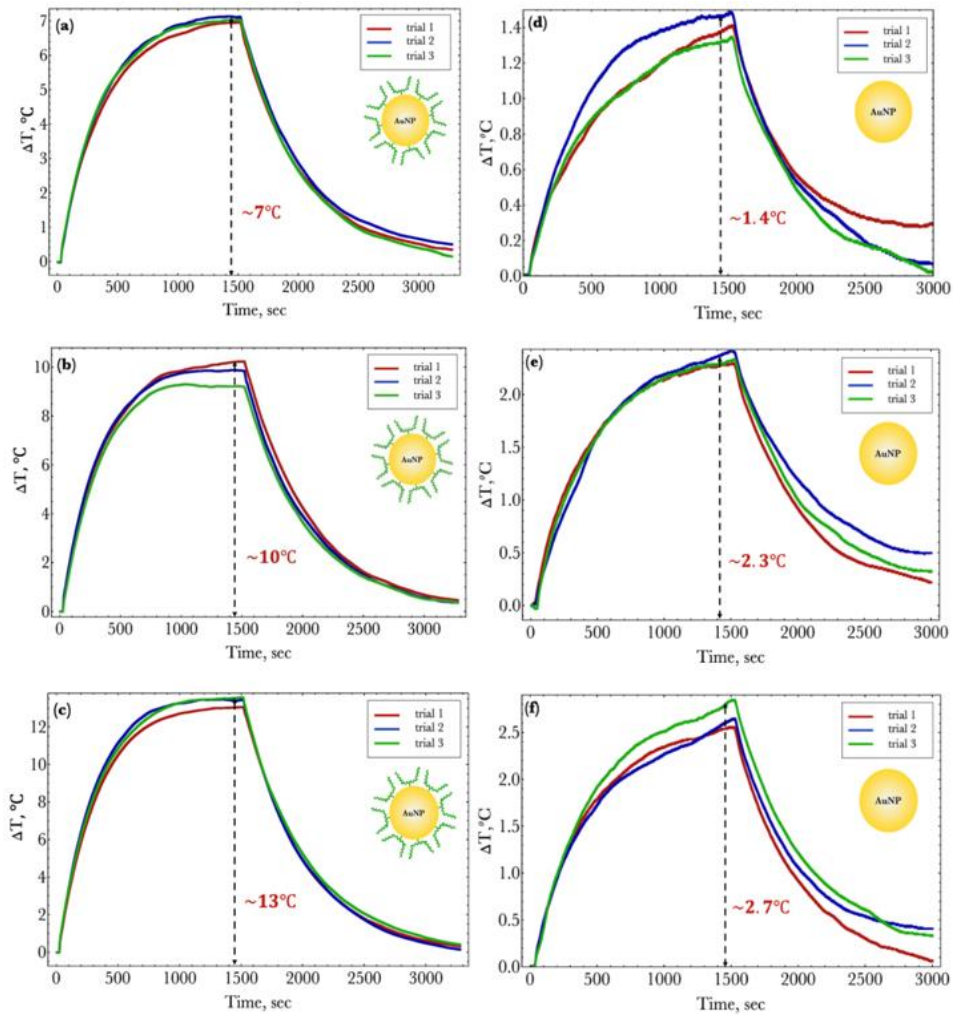

**Figure 2.** Temperature profiles for three trials for AuNP conjugates (a)-(c) and 20 nm AuNPs (d)-(f) in UPW for different laser powers: (a) and (d) Power is $0.5\,W$, (b) and (e) power density is $1\,W$, (c) and (f) power density is $2\,W$.

The maximum temperature of 13°C was reached for AuNP conjugates at the power of 2 $W$. For 1 $W$ and 0.5 $W$ powers, the $\Delta T = 10°C$ and $\Delta T = 7°C$ were recorded respectively (Fig. 2 (a)-(c)). The maximum temperature for the solution of 20 nm AuNPs was 2.7°C at power of 2 $W$ which in almost five times lower than temperature increase for the same power of laser in case of AuNP conjugates. $\Delta T = 1.4°C$ and $\Delta T = 2.3°C$ were reached for 20 nm AuNPs at 1 $W$ and 0.5 $W$ powers respectively. (see Supporting Info for temperature profiles of UPW for different NIR laser's powers)

AuNP conjugates absorb and scatter the incident light more efficiently as compared to 20 nm AuNPs. This is because of the presence of the second absorbance peak near the incident wavelength of the laser ($\lambda_2 = 780\ nm$) that is absent in gold nanospheres (Fig. 1(c)).

### 1.4   Photothermal Conversion Efficiency

The solution of gold nanoparticles, after being irradiated during some time, will eventually reach the thermal equilibrium when the temperature change is zero $\left(\frac{d\Delta T}{dt} = 0\right)$. Therefore, we can solve Eq. (5) with respect to the photothermal conversion efficiency:

$$\eta = \frac{B(T_{max}-T_0)}{(P_0-P_{tr})} m_W C_W. \tag{6}$$

The constant rate of heat dissipation $B$ from the gold nanoparticles to the external environment can be found from cooling down temperature profiles (Fig. 3). [35] Once the temperature of the solution reaches the saturation point, the NIR laser is turned off, which results in zero input heat $Q_{in} = 0$. The energy balance Eq. (5) then reduces to

$$\frac{d\Delta T}{dt} = -B\Delta T. \tag{7}$$

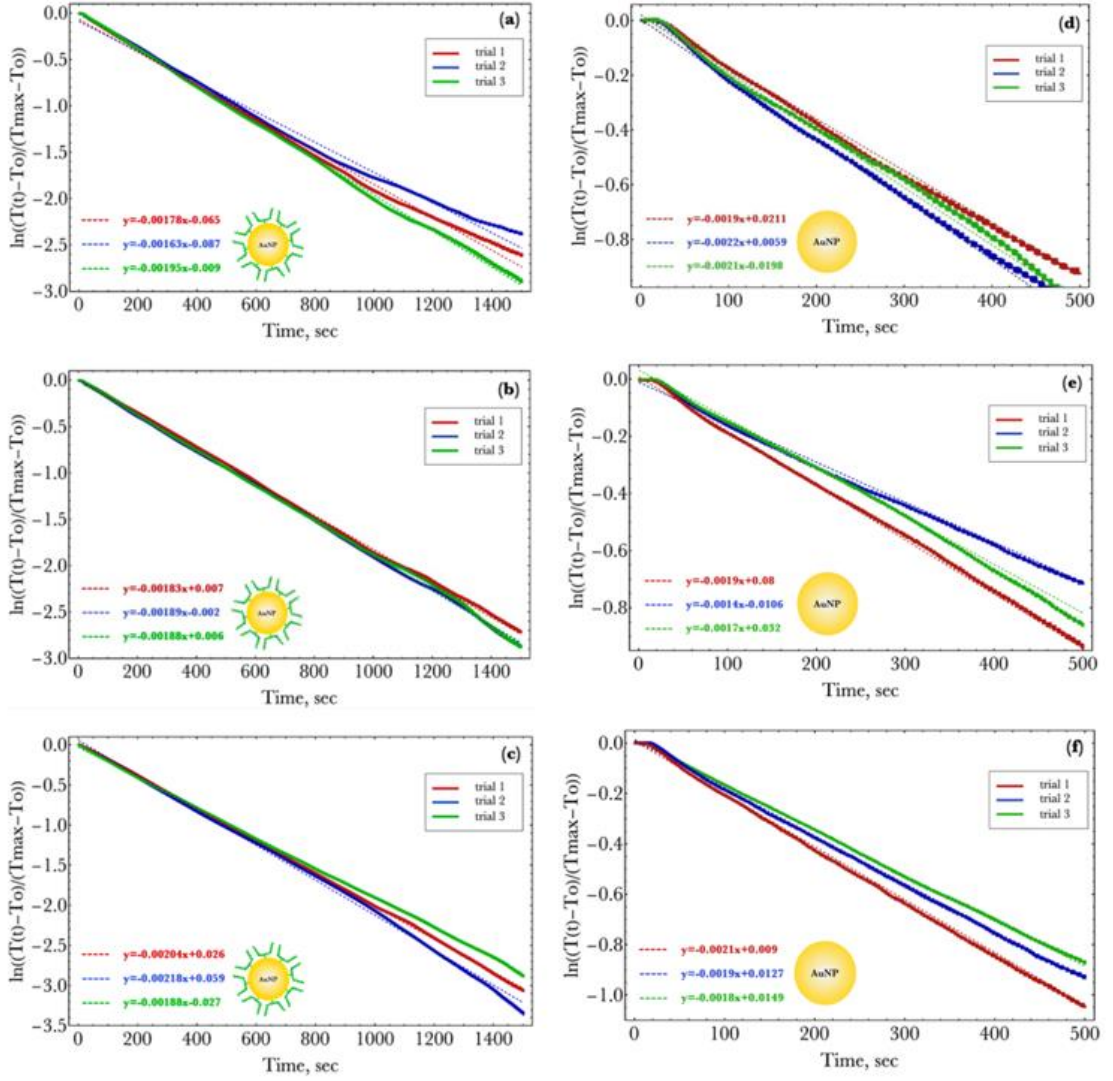

**Figure 3.** Cooling temperature profiles for three trials of AuNP conjugates (a)-(c) and 20 nm AuNPs (d)-(f) in UPW for different laser powers: (a) and (d) Power is $0.5\ W$, (b) and (e) power density is $1\ W$, (c) and (f) power density is $2\ W$.

The differential equation Eq. (7) can be solved to give the temperature profile of the sample

$$T(t) = T_0 + (T_{max} - T_0)\exp(-Bt), \tag{8}$$

where $T(0) = T_{max}$ is the maximum temperature when the laser is turned off. Note, Eq. (8) describes cooling down temperature profile. Rearranging Eq. (8) we find

$$ln\left(\frac{(T_{max}-T_0)}{(T(t)-T_0)}\right) = Bt. \tag{9}$$

Fig. 5 shows the $ln\left(\frac{(T_{max}-T_0)}{(T(t)-T_0)}\right)$ as a function of time $t$ for AuNP conjugates and 20 nm AuNPs. For each fitting of the curve, we find the slope of the equation that gives us the heat dissipation constant $B$. From Fig. 5 (a)-(c) one can find $B = 1.89 \pm 0.15 \ ms^{-1}$ for AuNP conjugates, and Fig. 5 (d)-(f) give us $B = 1.88 \pm 0.24 \ ms^{-1}$ for 20 nm AuNPs. Note, Fig. 5 (d)-(f) represents the cooling profile for the first 500 sec after the laser was turned off. 20 nm AuNPs don't heat significantly (Fig. 4 (d)-(f)), therefore the temperature equilibrium might be reached much faster than 25 minutes. After finding the $B$-values, we solve the Eq. (6) as

$$T_{max} - T_0 = \frac{\eta}{Bm_W C_W}(P_o - P_{tr}), \tag{10}$$

By plotting $\Delta T$ as a function of $(P_o - P_{tr})$, we can find the photothermal conversion efficiency for each solution. [35] The slope of Figure 6 represents the value of $\frac{\eta}{Bm_W C_W}$ expression.

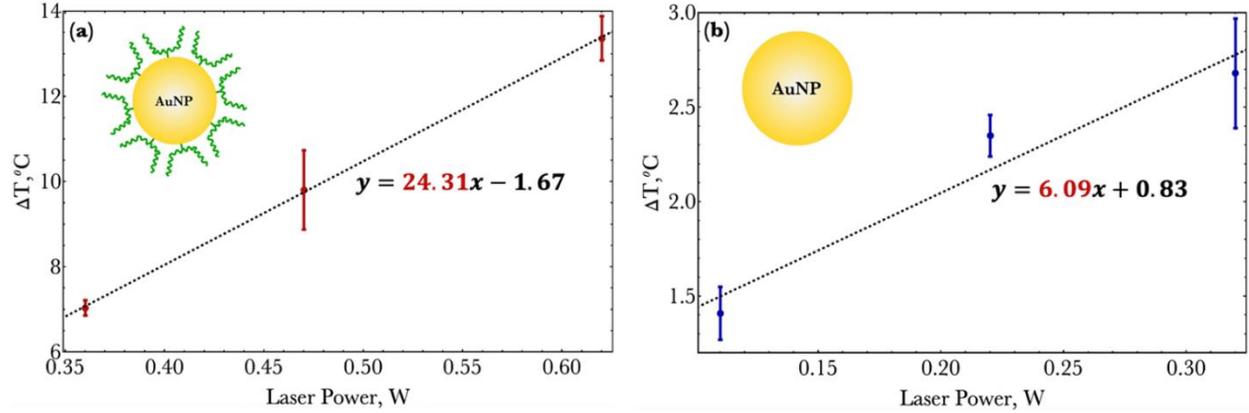

**Figure 4.** (a) $\Delta T$ vs $(P_o - P_{tr})$ graph for AuNP conjugates; (b) $\Delta T$ vs $(P_o - P_{tr})$ graph for 20 nm AuNPs. The error bars represent the range of the absolute value for $\Delta T$ for each solution based on measured triplicate (Fig. 2).

The best-fit line for $\Delta T$ vs $(P_o - P_{tr})$ graph for AuNP conjugates (Fig. 4 (a)) is $y = 24.31x - 1.67$. Thus, $\frac{\eta}{Bm_W C_W} = 24.31$, and $\eta = 47.7\ \%$. The curve-fit equation $\Delta T$ vs $(P_o - P_{tr})$ graph for 20 nm AuNPs (Fig. 4 (b)) is $y = 6.09x + 0.83$. Therefore, $\frac{\eta}{Bm_W C_W} = 6.09$, and $\eta = 12\ \%$.

**The photothermal conversion efficiency of AuNP conjugates is almost four times larger than the photothermal conversion efficiency of bare 20 nm AuNPs.** This result opens a path of using the lower concentrations of drugs that potentially reduces the unwanted accumulation of cancer drugs in healthy tissue.

It is noteworthy that the absorbance of AuNP conjugates changes drastically after just one trial of heating (see Supporting Info). Due to photo-bleaching of dye conjugated gold nanospheres during heating, a new solution of AuNP conjugates was prepared before each trial at the concentration noted in Figure 1 (c).

## 1.5 Synergistic Effect

We can hypothesize that higher photothermal conversion efficiency of gold spherical conjugates results due to the additive heating produced by 20 nm AuNPs and IR cyanine dyes irradiated by the laser. In this section, we experimentally test this hypothesis by heating solutions of 20 nm AuNPs and IR cyanine dyes separately. We experimentally show that the $\Delta T$s of each component are not additive, which leads us to discard the additive heating hypothesis. We observe a larger temperature increase for "spherical conjugates" (Gold covalently bound to the tumor-targeting dye) compared to 20 nm AuNPs and IR cyanine dye when they are added together.

Figure 5 (a) shows the absorbance spectrum of AuNP conjugates in UPW and solution of IR dye in the methanol (the IR cyanine dye is hydrophobic; therefore it cannot be diluted in water). During the synthesis of spherical conjugates, we cannot measure or calculate the numbers of IR cyanine dye molecules on the surface of a single gold nanosphere. Spherical conjugates and IR cyanine dyes represent the same absorption peaks at $\lambda_2 \sim 780\ nm$. The second absorption peak of AuNP conjugates is because of the presence of IR cyanine dye on the surface of the nanoparticle. By matching this peak, in the first approximation, we can assume the concentration of organic dyes is the same in each solution.

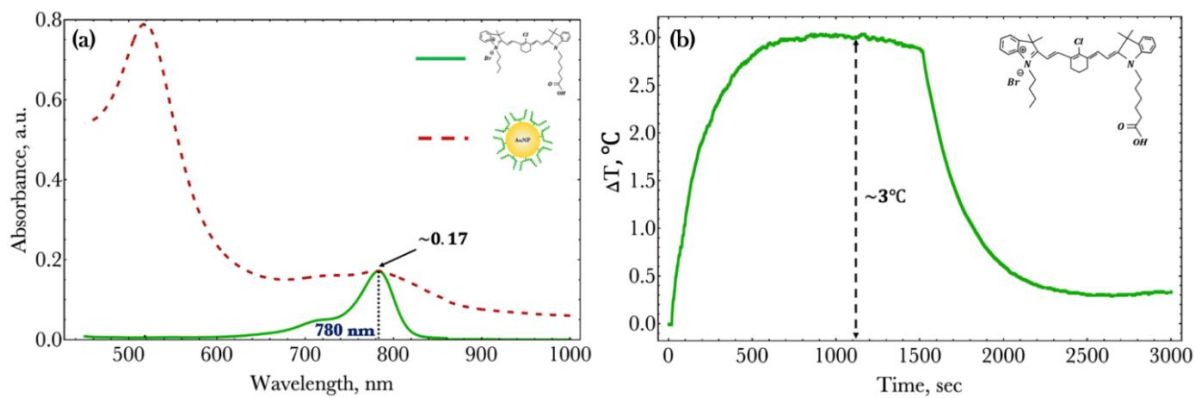

**Figure 5.** (a) Absorbance spectra of AuNP conjugates in UPW and cyanine IR dyes in methanol. (b) Temperature profile for cyanine IR dyes in methanol at laser's power of $1\,W$.

Figure 5 (b) depicts the temperature profile of IR cyanine dyes in methanol irradiated with the laser power of $1\,W$. The temperature increase for the solution of organic dyes is $3\,°C$, which is larger than $\Delta T$ for 20 nm AuNPs and is less than $\Delta T$ for AuNP conjugates (Figure 2). The temperature profile of the solution of IR cyanine dyes resembles the temperature profile of AuNP conjugates. Also, during the laser's irradiation, the organic dyes reach the saturation temperature relatively quickly, and after the first 20 min the local temperature in the solution decreases. This phenomenon is observed due to photobleaching of the dye. (see Supporting Info for temperature profiles of methanol for different NIR laser's powers)

By adding the temperature profile of IR cyanine dyes in methanol (Fig. 5 (a)) and the temperature profile of 20 nm AuNPs in water (Fig. 2 (d)), we obtain the total temperature profile of the "mixture" solution (20 AuNPs + IR cyanine dyes) that is shown on Figure 6. The maximum temperature increase is around $4.5\,°C$. The temperature increases for AuNP conjugates, irradiated by the NIR laser light at the power of $1\,W$, is $7\,°C$ (Fig. 2 (a) and Fig. 6) which is much larger than $\Delta T$ for gold nanospheres plus dyes solution. Gold nanospheres conjugated with infrared dyes exhibit synergistic effects that result in higher photothermal conversion efficiency.

By conjugating gold nanospheres with IR cyanine dyes, one can induce the coupling between the metal nanoparticle and the organic emitter that leads to the dominance of the nonradiative decay rate over the radiative decay channel [36-38].

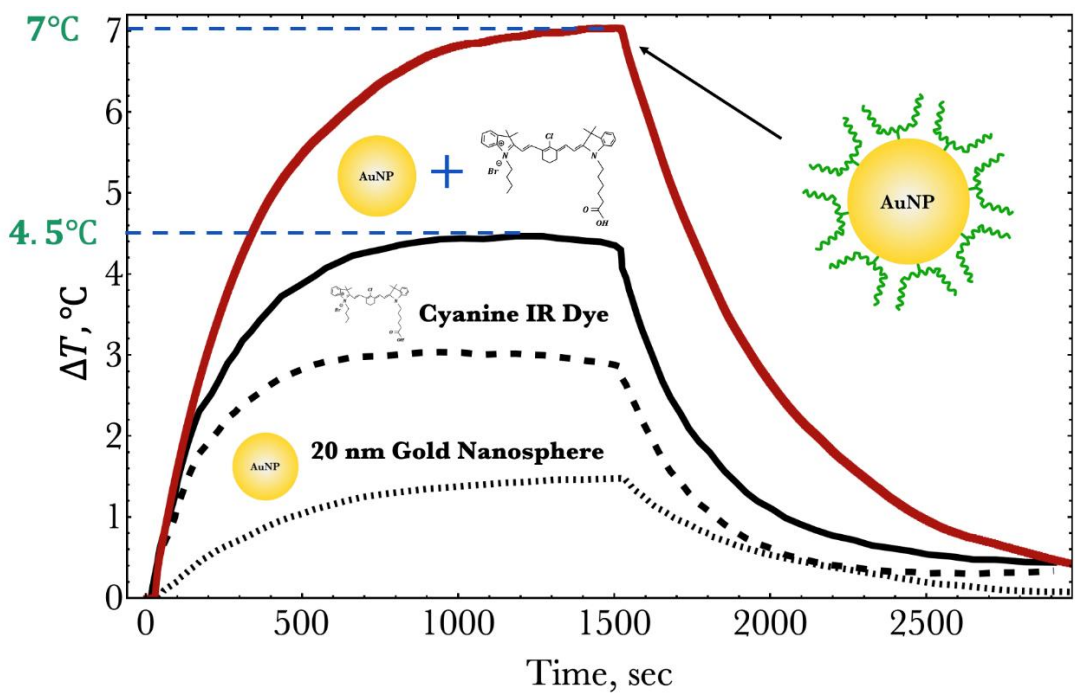

**Figure 9.** Temperature profile for AuNP conjugates in UPW, 20 nm AuNPs in UPW, IR cyanine dyes in methanol, and for 20 nm AuNPs + IR cyanine dye mixture at laser's power of $1\ W$

## 1.6    Conclusions

Gold nanospheres conjugated with tumor-targeting organic dyes have enhanced photothermal conversion efficiency as compared to bare spherical gold nanoparticles. The photothermal conversion efficiency of spherical conjugates is almost four times larger than the photothermal efficiency of gold nanospheres. We experimentally observed this strong and repeatable synergistic heating effect during the laser irradiation process. We experimentally showed that by adding temperature profiles for 20 nm AuNPs and IR cyanine dyes, we don't get the same temperature increase that we observe for spherical gold conjugates. These results suggest that there is particle-dye coupling leading to the observed synergistic effect.

Conjugated metal nanoparticles of any shape and size with the fluorescent tumor-targeting dyes can be used as a novel drug for theranostic purposes. It is a promising platform for non-invasive treatment of a variety of cancers. In addition to the imaging modality of the dyes themselves and active targeting of the cancer tissues, the synergistic effect described above can be very beneficial for photothermal therapy. One may achieve sufficient temperatures to treat the cancer cells at a greatly reduced concentration of drug and/or lower power of the laser. It is of great interest to study a gold nanoparticle with the LSPR in the NIR region synthesized with near-infrared dye. It's expected that such structures may produce even larger photothermal properties than discussed in this paper.